\documentclass[12pt]{iopart}
\usepackage{rotating}
\usepackage{rotate}
\usepackage{epsf}
\usepackage{pifont}
\usepackage{latexsym}
\usepackage{epsfig}
\usepackage{rotating}
\usepackage{graphicx}
\usepackage{float}
\usepackage{amssymb}
\usepackage{wrapfig}

\newcommand{\dd}{{\rm d}}
\newcommand{\sqrtsnn}{\sqrt{s_{_{{\rm N N}}}}}
\def\z{z_{_{\gamma \pi}}}
\def\cO#1{{{\cal{O}}}\left(#1\right)}
\def\picut{p_{_{\perp_\pi}}^{\rm cut}}
\def\gacut{p_{_{\perp_\gamma}}^{\rm cut}}
\def\ptpi{p_{_{\perp_\pi}}}
\def\ptgamma{p_{_{\perp_\gamma}}}
\def\lQCD{\Lambda_{_{\rm QCD}}}
\def\kt{{k_{_\perp}}}
\def\pt{{p_{_\perp}}}
\def\gampi{\gamma$--$\pi^0} 

\begin{document}

\begin{flushright}
CERN-PH-TH/2007-005; LAPTH-1173/07; \texttt{hep-ph/0701207}
\end{flushright}

\title[Photon-tagged correlations: kinematic requirements and a case study]{Photon-tagged correlations in heavy-ion collisions: kinematic requirements and a case study\footnote{Based on talks given at Quark Matter 2006, Shanghai, China, 14-20 November 2006, and at the ALICE Physics Working Group~4, CERN, 13 December 2006.}}

\author{Fran\c{c}ois Arleo \footnote{On leave from Laboratoire d'Annecy-le-Vieux de Physique Th\'eorique (LAPTH), UMR 5108 du CNRS associ\'ee \`a l'Universit\'e de Savoie, B.P. 110, 74941 Annecy-le-Vieux Cedex, France.}}

\address{CERN, PH Department, TH Division, 1211 Geneva 23, Switzerland}
\ead{arleo@cern.ch}

\begin{abstract}
Photon-tagged correlations may be useful to determine how the dense partonic medium produced in heavy-ion collisions affects the fragmentation of high-energy quarks and gluons into a leading hadron. In these proceedings, I discuss the kinematic requirements for the hadron and the prompt photon transverse momentum cuts. A case study at LHC energy, tagging on $\ptgamma\ge{20}$~GeV and $\ptgamma\ge{50}$~GeV photons, is then briefly examined.
\end{abstract}

\section{Why photon-tagged correlations}
\label{sec:introduction}

\begin{wrapfigure}[10]{r}{7.cm}
\vspace{-0.9cm}
  \begin{center}
    \includegraphics[width=7.cm]{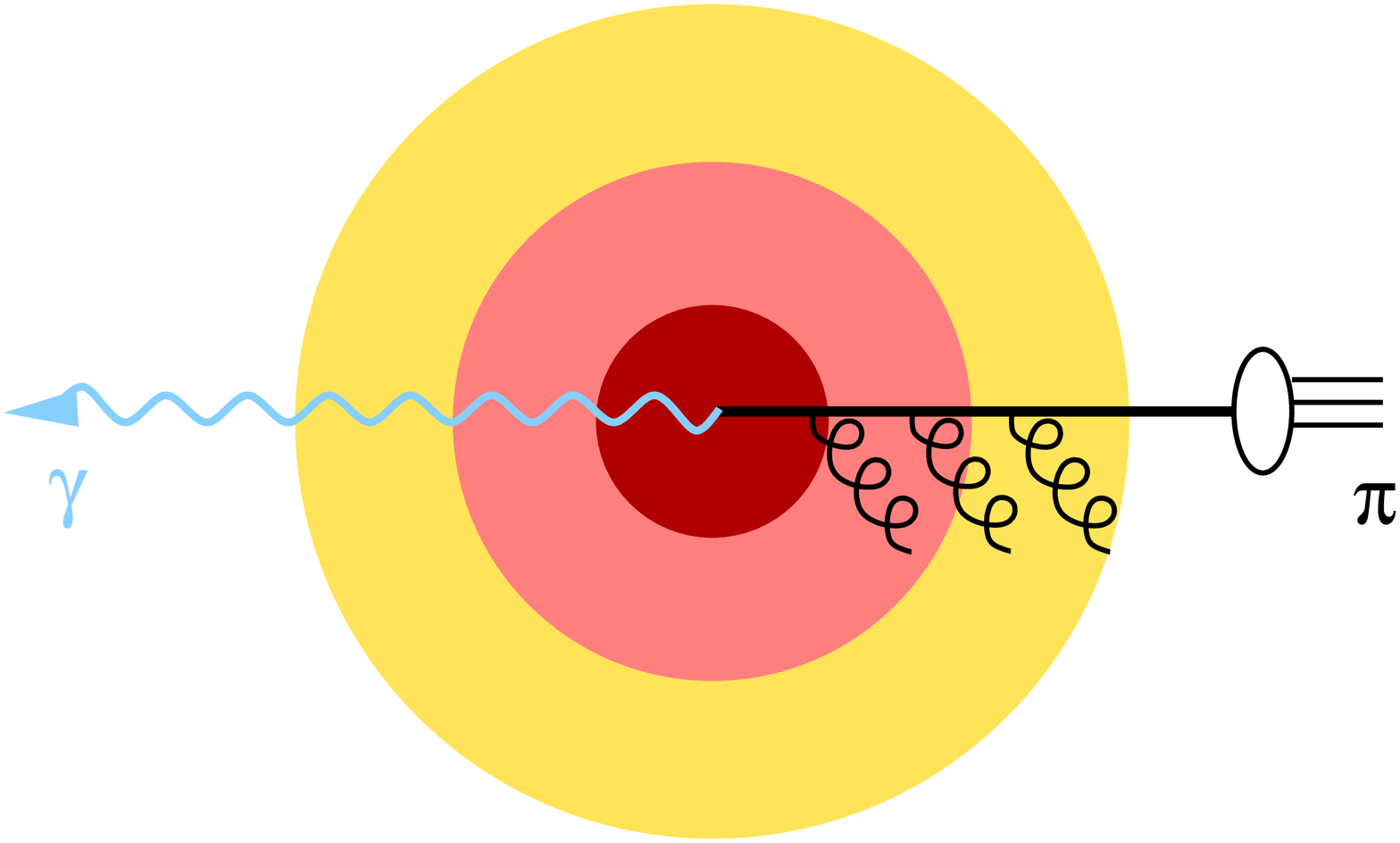}
\vspace{-0.8cm}
    \caption{A leading-order channel for $\gampi$ production.}
  \end{center}
\label{fig:sketch}
\end{wrapfigure}

The spectacular quenching of large $\pt$ pion production measured in central Au--Au over that in $p$--$p$ reactions at RHIC is an important discovery, clearly indicating the formation of a dense partonic system in the early stage of these heavy-ion collisions~\cite{Adcox:2004mhAdams:2005dq}. However, this observation poorly informs us on {\it how} exactly the produced medium affects the propagation and the hadronization of high-energy quarks and gluons. Indeed, unlike jets in $e^+e^-$ or deep inelastic scattering events, the single pion hadroproduction spectra do not allow the initial parton momentum $\kt$ --~and therefore the fragmentation variable $z = \ptpi / \kt$~-- to be determined. 
One way to possibly access medium-modified fragmentation functions in heavy-ion collisions would be to tag large-$\pt$ particle production with prompt photons, as suggested in~\cite{Wang:1996yh} and later investigated in detail at RHIC and LHC energy~\cite{Arleo:2004xjArleo:2006xb}. Take for instance the process sketched in Figure~\ref{fig:sketch}: a photon is produced directly in the hard subprocess, back-to-back to a quark which loses some energy (through medium-induced gluon radiation) before eventually fragmenting into a leading pion. Because of momentum conservation, the $\gampi$ momentum imbalance variable,
\begin{equation}
\z \equiv - \frac{{\bf \ptpi} . {\bf \ptgamma}}{|{\bf \ptgamma}|^2},
\end{equation}
reduces to the fragmentation variable, $\z = z$, in this leading-order (LO) kinematics. Therefore, there should be a clear connection between the {\it experimentally} accessible momentum-imbalance distributions and the {\it theoretical} quark fragmentation function into a pion: 
\begin{equation}
  \label{eq:correspondence}
  {\rm {\footnotesize (experimental)}} \qquad \frac{\dd\sigma}{\dd\z} \quad \Leftrightarrow \quad D_q^\pi(z) \qquad {\rm (theoretical)}
\end{equation}

\noindent This reasoning is of course too simplistic, since the photon can itself be produced by the collinear fragmentation of a leading parton. In this case, there is no correlation between the pion and the photon momenta, and the correspondence (\ref{eq:correspondence}) is lost. As we shall see, this is precisely the most important ``background'' channel we would like to reduce with appropriate kinematic cuts. Moreover, higher-order corrections as well as initial- and final-state soft gluon radiation could complicate somehow the picture.

I will first specify in Section~\ref{sec:requirements} under which kinematic conditions the $\gampi$ momentum-imbalance distributions may reflect the quark fragmentation function into the pion. A simple case study at the LHC is then discussed in Section~\ref{sec:casestudy}.

\section{Kinematic requirements}
\label{sec:requirements}

(i) \underline{A truly perturbative pion}\\

\noindent In order to construct meaningful $\gampi$ correlations, the pion needs to be produced perturbatively --~that is coming from the fragmentation of a high-energy quark/gluon~-- therefore with momentum $\ptpi \gg \lQCD=\cO{1\ {\rm GeV}}$. On top of this, $\ptpi$ should ideally be much larger than the typical scales of the medium, in order to safely exclude new production mechanisms such as recombination, which may take place at RHIC in the $\ptpi\simeq 2$--$5$~GeV range~\cite{Fries:2003vb}. Because of the more dramatic environment at the LHC (higher energy density, stronger flow, \dots), it could be wise to require the pion momentum cut, $\picut$, to be greater than 10~GeV at the LHC.\\

\noindent (ii) \underline{A wide $\z$ range}\\

\noindent If there is any matching between imbalance distributions and $D_q^\pi(z)$, the window at which this occurs should of course be as large as possible, i.e. $z_{_{\rm min}} \lesssim \z \lesssim 1$, with $z_{_{\rm min}}\ll 1$. Suppose that the photon momentum is fixed to $\ptgamma=\gacut$, the corresponding pion momentum should vary from $\ptpi\simeq z_{_{\rm min}}\ \ptgamma \ll \gacut$ ($\z \simeq z_{_{\rm min}}$) up to $\ptpi\simeq \ptgamma$ ($\z \simeq 1$). Therefore, the $\z$ range over which a matching with the fragmentation function could be expected is $\picut/\gacut \lesssim \z \lesssim 1$. In other words, the more asymmetric the pion and the photon $\pt$ cuts, the better\footnote{As a matter of fact, the LO $\z$ distribution has a natural maximum at the edge of the selected momentum-space, $\ptpi\gtrsim\picut$ and $\ptgamma\gtrsim \gacut$, that is for pairs with $\z\simeq \picut/\gacut$. Lower $\z$ could only be reached by {\it increasing} the photon momentum (since $\ptpi$ cannot be lowered), thus with a lower cross section. This is at variance with $D_q^\pi(z)$ which monotonically decreases with $z$.}.\\

\noindent (iii) \underline{A direct photon}\\

\noindent As already emphasized, the prompt photon needs to be produced ``directly'' in the partonic process (Figure~\ref{fig:sketch}) in order to reconstruct fragmentation functions\footnote{The most promising way to get rid of most of the photons coming from quark fragmentation is to apply isolation criteria, although this may be difficult in heavy-ion collisions with high multiplicity.}. This channel should be dominant at large $x=2\ \ptgamma/\sqrtsnn\ll{1}$, when the momentum-space to pick up higher energy  $\kt>\ptgamma$ partons in the nucleons is restricted. Above $x\simeq{10^{-2}}$, the gluon distributions are strongly suppressed, and the direct process is favoured. At the LHC ($\sqrtsnn=5.5$~TeV in Pb--Pb), this translates into $\gacut\gtrsim{30}$~GeV.\\

\noindent (iv) \underline{Reasonable counting rates}\\

\noindent Last but not least, the number of events should remain large enough so that the imbalance distributions can be measured with a high-enough statistical accuracy. The photon momentum cut hence needs to be much smaller than the kinematic boundary, $\gacut \ll \sqrtsnn$. Its precise value depends of course crucially on the integrated luminosity; as we shall see in the next Section, $\gacut$ has to be strictly below $100$~GeV at the LHC.

\section{Case study at the LHC}
\label{sec:casestudy}

In order to illustrate the above requirements, the $\gampi$ imbalance distributions are computed perturbatively in Pb--Pb collisions at the LHC. The calculations are carried out to LO in $\alpha_s$, using the standard fragmentation functions in $p$--$p$ reactions and the rescaled fragmentation functions in central Pb--Pb collisions~\cite{Wang:1996yh} using the BDMPS quenching weight~\cite{Arleo:2002kh} (see~\cite{Arleo:2004xjArleo:2006xb} for details). The following cuts are chosen:
\begin{center}
  {\bf Case A:} $\picut=10$~GeV $\bigr/$ $\gacut=20$~GeV \hspace{0.4cm}{\bf Case B:} $\picut=10$~GeV $\bigr/$ $\gacut=50$~GeV
\end{center}
Clearly, Case A has the main advantage to offer higher rates than Case B. On the contrary, the $\pi/\gamma$ cuts are not too asymmetric ($\picut/\gacut=1/2$), and the photons could be abundantly produced through fragmentation because of the not too large cut, $\gacut=20$~GeV.

\begin{figure}[htb]
\begin{center}
\includegraphics[width=6.2cm]{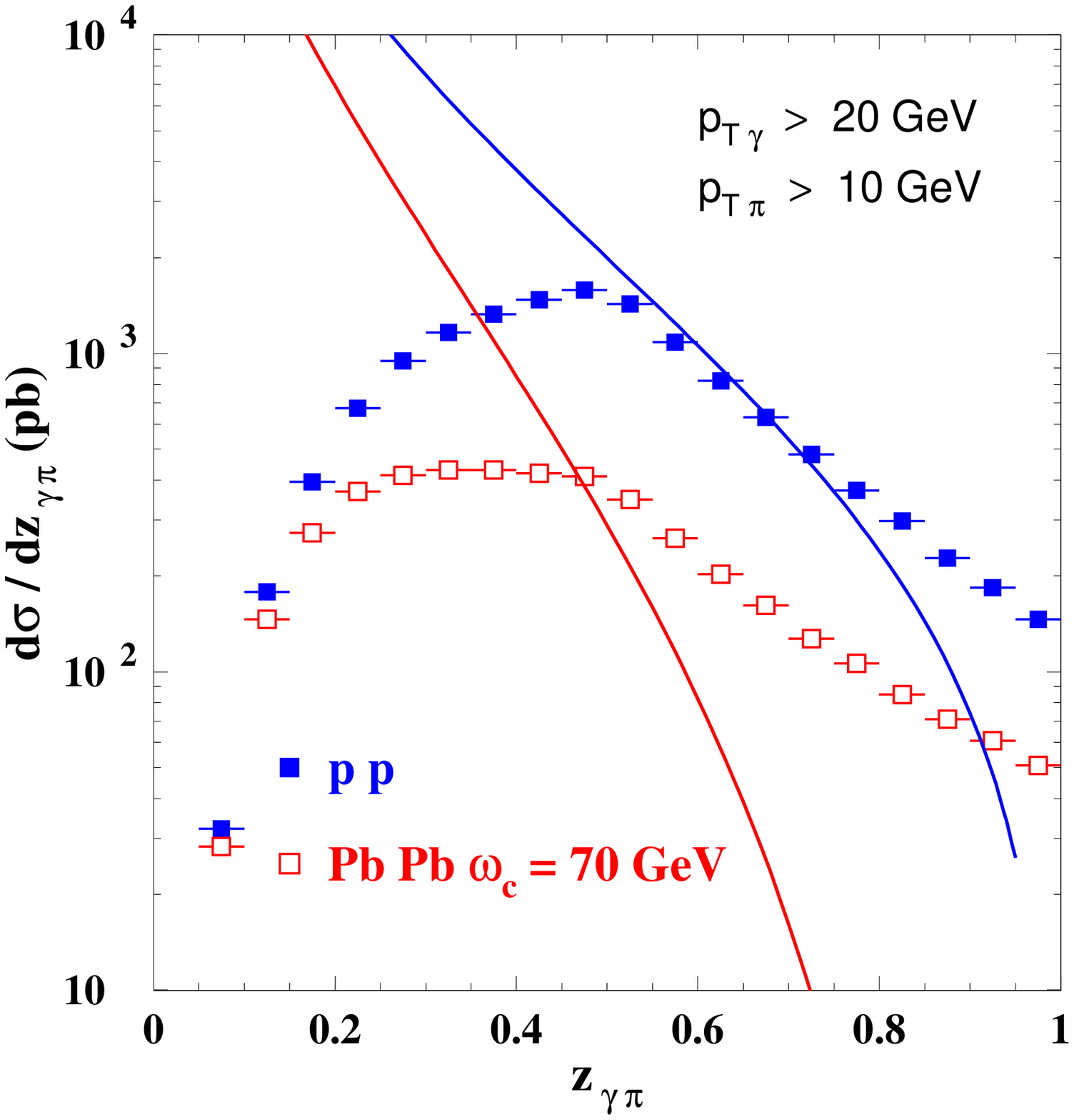}
\includegraphics[width=6.2cm]{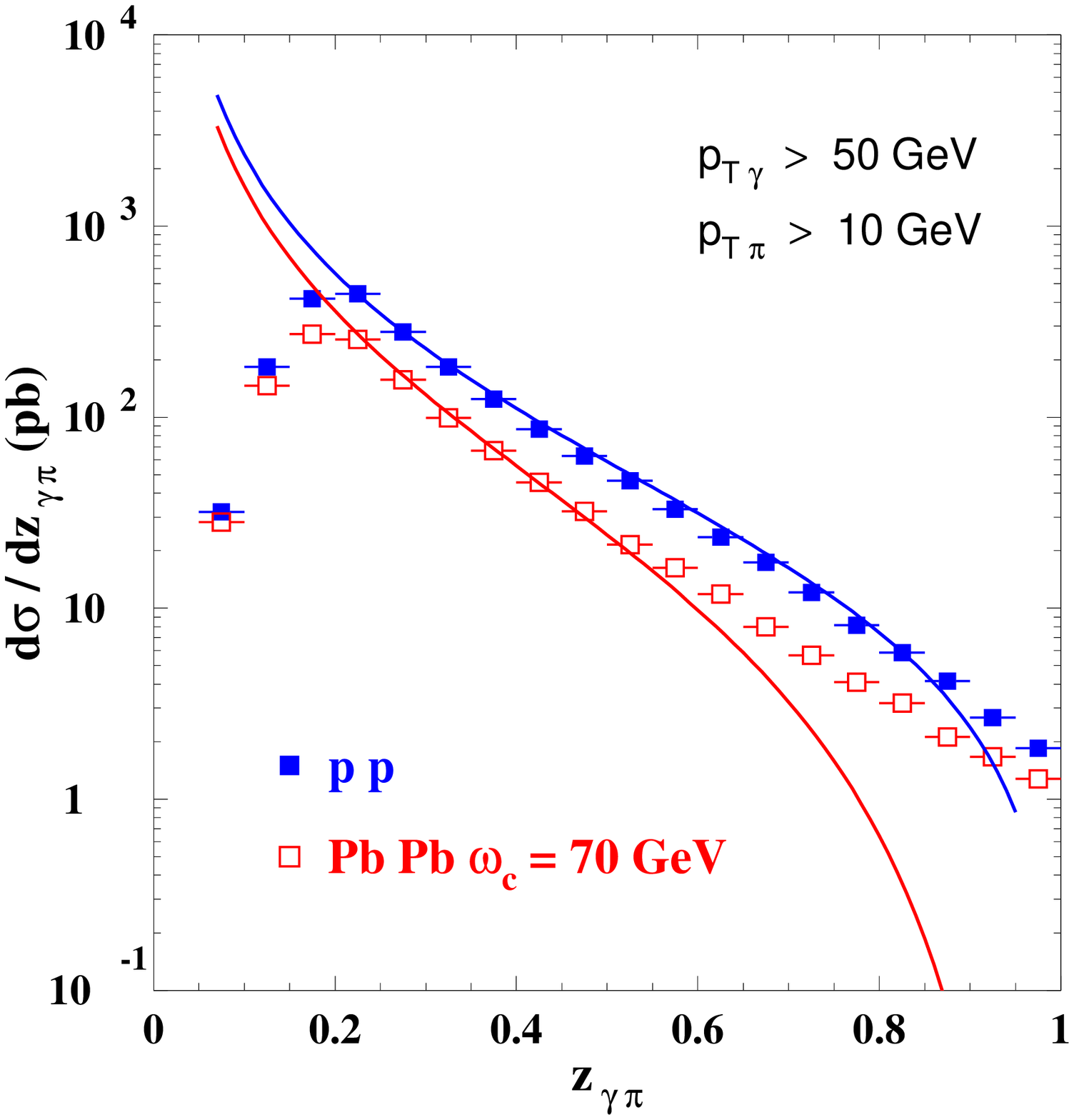}
\end{center}
\vspace{-0.6cm}
\caption{$\gampi$ imbalance distributions in Case A (left) and B (right). The solid lines show the (rescaled) quark fragmentation functions used in the calculation.}
\label{fig:distributions}
\end{figure}

The $\gampi$ imbalance distributions in $p$--$p$ and Pb--Pb collisions are plotted in Figure~\ref{fig:distributions} for the Case A (Left) and B (right). As anticipated, these distributions show a maximum at $\picut/\gacut$. In order to show the possible connection between $\dd\sigma/\dd\z$ and $D_q^\pi(z)$, the ``input'' vacuum and medium-modified quark fragmentation into the pion used in the calculation (arbitrarily rescaled so as to match $p$--$p$) are also displayed. When $\gacut=20$~GeV, there is absolutely no matching between those two quantities. This is because of the important contribution of the fragmentation-photon channel, as well as the restricted kinematical $\z$ window, $
0.5 \le \z \le 1$, over which imbalance distributions and fragmentation functions could have matched. On the contrary, a much better ``agreement'' is found in Case B, from $\z\simeq \picut/\gacut=0.2$ up to $\z\simeq 0.9$ in $p$--$p$ and $\z\simeq 0.6$ in Pb--Pb collisions, over which the fragmentation-photon channel becomes predominant.
The counting rates for the most central ${\cal C}\le 20\%$ Pb-Pb collisions in the ALICE detector, using the year-integrated luminosity ${\cal L}_{_{\rm int}} = 5.10^{-4}\,{\rm pb}^{-1}$, are now given. In Case A, $\dd\sigma/\dd\z$ roughly ranges from $5.10^1$ to $5.10^2$~pb, which translates into $\dd{\cal N}/\dd\z \sim{7.10^2}$~--~$7.10^3$ events (see~\cite{Arleo:2003gn}). Cross sections are of course somewhat smaller in case B, $\dd\sigma/\dd\z\sim{1}$~--~$3.10^2$~pb, corresponding to $\dd{\cal N}/\dd\z\sim{15}$--$4.10^3$. Despite the smaller number of events, it seems that $\gacut=50$~GeV proves by far more advantageous than the lower cut $\gacut=20$~GeV, which clearly misses the main interest of photon-tagged correlations: probing vacuum and medium-modified fragmentation functions.

In summary, a clear hierarchy among the different energy scales involved needs to be respected: $\lQCD, T \ll \picut \ll \gacut \ll \sqrtsnn$. At LHC, $\picut$ should be of the order of 10~GeV, while $\gacut$ definitely needs to be at least 50~GeV in order to fully appreciate the potential richness of photon-tagged correlations in heavy-ion collisions.

\vspace{-0.4cm}
\section*{References}
\providecommand{\href}[2]{#2}\begingroup\raggedright
\endgroup


\begin{thebibliography}{10}
\bibitem{Adcox:2004mhAdams:2005dq}
Adcox K et~al., Nucl. Phys. {\bf A757} (2005) 184--283 [\href{http://arXiv.org/abs/nucl-ex/0410003}{{\tt nucl-ex/0410003}}];\\
\hspace{-0.45cm} Adams J et~al., Nucl. Phys. {\bf A757} (2005) 102--183 [\href{http://arXiv.org/abs/nucl-ex/0501009}{{\tt nucl-ex/0501009}}].

\bibitem{Wang:1996yh}
Wang X-N, Huang Z and Sarcevic I, Phys. Rev. Lett. {\bf 77} (1996) 231--234
  [\href{http://arXiv.org/abs/hep-ph/9605213}{{\tt hep-ph/9605213}}].

\bibitem{Arleo:2004xjArleo:2006xb}
Arleo F, Aurenche P, Belghobsi Z and Guillet J-P, JHEP {\bf 11} (2004) 009
  [\href{http://arXiv.org/abs/hep-ph/0410088}{{\tt hep-ph/0410088}}];\\
\hspace{-0.45cm} Arleo F, JHEP {\bf 09} (2006) 015 [\href{http://arXiv.org/abs/hep-ph/0601075}{{\tt hep-ph/0601075}}].

\bibitem{Fries:2003vb}
Fries R J, M\"uller B, Nonaka C and Bass S A, Phys. Rev. Lett. {\bf 90} (2003) 202303 [\href{http://arXiv.org/abs/nucl-th/0301087}{{\tt nucl-th/0301087}}].

\bibitem{Arleo:2002kh}
Arleo F, JHEP {\bf 11} (2002) 044 [\href{http://arXiv.org/abs/hep-ph/0210104}{{\tt hep-ph/0210104}}].

\bibitem{Arleo:2003gn}
Arleo F et~al., ``Photon physics'' section, CERN Yellow Report 2004-009 [\href{http://arXiv.org/abs/hep-ph/0311131}{{\tt hep-ph/0311131}}].

\end{thebibliography}
\end{document}